\documentclass[rnote]{aa}
\usepackage[varg]{txfonts}

\usepackage{natbib}
\bibpunct{(}{)}{;}{a}{}{,} 
\usepackage{txfonts}

\newcommand{\ctan}{$^{13}$C($\alpha$,n)$^{16}$O }
\newcommand{\nean}{$^{22}$Ne($\alpha$,n)$^{25}$Mg }
\newcommand{\ct}{$^{13}$C }
\newcommand{\fluod}{$^{19}$F }

\begin{document}

\title{Effects of nuclear cross sections on $^{19}$F nucleosynthesis at low metallicities}

   \author{S.~Cristallo
          \inst{1,2}
          \and
           A.~Di Leva
          \inst{3,2}
          \and
          G.~Imbriani
          \inst{3,2}
          \and
          L.~Piersanti
          \inst{1,2}
          \and
          C.~Abia
          \inst{4}
          \and
          L.~Gialanella
          \inst{5,2}
          \and
          O.~Straniero
          \inst{1,2}
          }

   \offprints{S. Cristallo}

      \institute{
              INAF, Osservatorio Astronomico di Collurania, 64100 Teramo, Italy\\
              \email{sergio.cristallo@inaf.it}
                            \and
                            INFN Sezione Napoli, Napoli (Italy)\\
                     \and
                     Dipartimento di Fisica, Universit\`a di Napoli ``Federico II'', Napoli,
                     Italy\\
                     \and
                     Departamento de Fisica Teorica y del Cosmos, Universidad de Granada, 18071 Granada, Spain\\
                     \and
                     Dipartimento di Matematica e Fisica, Seconda Universit\`a di Napoli, Caserta,
                     Italy\\
             }

   \date{Received ; accepted }

  \abstract
   {
   The origin of fluorine is a longstanding problem in nuclear
astrophysics. It is widely recognized that Asymptotic Giant Branch
(AGB) stars are among the most important contributors to the
Galactic fluorine production.
   }
   {
In general, extant nucleosynthesis models overestimate the
fluorine production by AGB stars with respect to observations.
Although at solar metallicity those differences are rather small,
low metallicity AGB stellar models predict fluorine surface
abundances up to one order of magnitude larger than the observed
ones. }
   {
As part of a project devoted to the reduction of the nuclear
physics uncertainties affecting the nucleosynthesis in AGB stellar
models, in this paper we review the relevant nuclear reaction
rates involved in the fluorine production/destruction. We perform
this analysis on a model with initial mass M=2 M$_\odot$ and
Z=0.001~.}
   {
We found that the major uncertainties are due to the
$^{13}$C($\alpha$,n)$^{16}$O, the $^{19}$F($\alpha$,p)$^{22}$Ne
and the $^{14}$N(p,$\gamma$)$^{15}$O reactions. A change of the
corresponding reaction rates within the present experimental
uncertainties implies surface $^{19}$F variations at the AGB tip
lower than 10\%, thus much smaller than observational
uncertainties. For some $\alpha$ capture reactions, however, cross
sections at astrophysically relevant energies are determined on
the basis of nuclear models, in which some low energy resonance
parameters are very poorly known. Thus, larger variations in the
rates of those processes cannot be excluded. That being so, we
explore the effects of the variation of some $\alpha$ capture
rates well beyond the current published uncertainties. The largest
$^{19}$F variations are obtained by varying the
$^{15}$N($\alpha$,$\gamma$)$^{19}$F and the
$^{19}$F($\alpha$,p)$^{22}$Ne reactions.}
{The currently estimated uncertainties of the nuclear reaction
rates involved in the production and destruction of fluorine
produce minor $^{19}$F variations in the ejecta of AGB stars. The
analysis of some $\alpha$ capture processes assuming a wider
uncertainty range determines $^{19}$F abundances in better
agreement with recent spectroscopic fluorine measurements at low
metallicity. In the framework of the latter scenario the
$^{15}$N($\alpha$,$\gamma$)$^{19}$F and the
$^{19}$F($\alpha$,p)$^{22}$Ne reactions show the largest effects
on fluorine nucleosynthesis. The presence of poorly known low
energy resonances make such a scenario, even if unlikely,
possible. We plan to directly measure these resonances.}

   \keywords{Physical data and processes: nuclear reaction,
   nucleosynthesis, abundances; stars: AGB and post-AGB}

   \authorrunning{Cristallo et al.}
   \titlerunning{$^{19}$F nucleosynthesis at low metallicities}
   \maketitle

\section{Introduction} \label{intr}

The production of fluorine ($^{19}$F) is an intriguing and largely
debated problems in stellar nucleosynthesis. Several stellar sites
have been proposed as $^{19}$F factories: core-collapse Supernovae
\citep{wax88}, Wolf-Rayet stars \citep{mey00} and Asymptotic Giant
Branch (AGB) stars \citep{forestini92}. While some papers excluded
the first two scenarios \citep{federman05,palacios05}, more recent
studies stressed the importance of both SuperNovae \citep{koba11}
and Wolf-Rayet stars \citep{jo14}. Among all the proposed
scenarios, however, a direct proof of fluorine production has been
proved only in AGB stars, via spectroscopic detections of [F/Fe]
enhancements (see \citealt{abia09} and references therein).

The first systematic search of fluorine enhancements in AGB stars
was done by \cite{jori92}, who measured abundances for red giants
of type K, Ba, M, MS, S, SC, N and J of near solar metallicity.
The result of this study was very high $^{19}$F surface
enrichments (up to 30 times solar) in N type C-stars and a clear
correlation between the fluorine enhancement and the C/O ratio.
N-type C stars are low-mass stars close to the AGB tip. Although
this occurrence was immediately interpreted as clear evidence of
the fluorine synthesis by AGB stars, theoretical models failed in
reproducing such a large surface fluorine enrichment
\citep{forestini92,lugaro04,cri09}. However, a re-analysis made by
\cite{abia09,abia10} of the same sample reconciled theoretical
models with observations. A systematic reduction of fluorine
abundances, by 0.7 dex on average, was found, relating this
discrepancy to blends with C-bearing molecular lines not properly
taken into account in \cite{jori92}. As a matter of fact, at
solar-like metallicities, the agreement between theory and
observations is rather good. However, extant fluorine
observational data may suffer a non negligible reduction due to an
error in the adopted HF excitation energies of infrared lines (see
\citealt{jo14}). This might systematically diminish the
spectroscopic $^{19}$F
estimates by 0.1 dex up to 0.5 dex in the cooler AGB stars.\\
At low metallicities, the situation is even more complex. Fluorine
measurements in three extragalactic low metallicity
(-2$<$[Fe/H]$<$-1) AGB stars \citep{abia11} disagree with
theoretical models, which predict fluorine surface abundances
about an order of magnitude larger (see \citealt{cri09,cris11},
and also Figures 3 and 4 in \citealt{abia11}). On the other hand,
these objects are largely enriched in s-process elements, with
0.9$<$[$\langle$s$\rangle$/Fe]\footnote{[$\rm\langle
s\rangle$/Fe]=0.5*(([Sr/Fe]+[Y/Fe]+[Zr/Fe])/3+
([Ba/Fe]+[La/Fe]+[Nd/Fe]+[Sm/Fe])/4). The two addends represent
the surface enrichment of elements belonging to the first peak and
second peak of the s-process, respectively.}$<$1.6
\citep{dela06,abia08}, thus indicating that the s-process
nucleosynthesis is connected to the fluorine production in AGB
stars. This implies quite low [F/$\rm\langle s\rangle$] values.
\noindent \citet{luca11} also found low [F/$\rm\langle s\rangle$]
in a sample of Galactic s-process rich carbon-enhanced metal-poor
stars, thus confirming
the results of \cite{abia11}.  \\
AGB stars are among the most important polluters of the
inter-stellar medium. In fact, they eject both light (C, N, F, Na)
and heavy elements (see \citealt{ir83,stra06} for a review). This
is the result of the combined action of internal nuclear burning
and deep convective mixing episodes taking place during the late
part of the AGB, the so-called Thermally Pulsing AGB or TP-AGB
phase. The presence of free neutrons in the He-rich intershell of
these stars, which is required to synthesize elements heavier than
iron, also affects the light elements nucleosynthesis and, among
the
others, fluorine. \\
The major source of neutrons in low mass AGB stars is the \ctan~
reaction \citep{stra95,ga98}, which releases neutrons in the so
called $^{13}$C-pocket. It is clear from the '60thies that the \ct
left in the H-shell ashes is not sufficient to account for the
observed s-process abundances \citep{fo55}. Thus, the penetration
of some protons from the convective envelope into the underlying
radiative He-intershell is required during TDU episodes. Later,
when the temperature reaches $10^8$ K, those protons are captured
by the abundant $^{12}$C leading to the formation of a thin
$^{13}$C-enriched layer, the so-called $^{13}$C pocket. Actually,
the mechanism leading to the formation of the \ct pocket is far
from being completely understood. In the last decades, many
theories have been proposed (e.g. \citealt{sac74,iben82,ir83});
most recently, mechanisms as mechanical overshoot \citep{he97},
gravity waves \citep{deto03}, opacity-induced overshoot
\citep[][see Section
\ref{stema}]{stra06} and magnetic fields \citep{trip14} have been invoked. \\
An additional neutron burst is powered by the \nean~ reaction,
which is activated at the base of the convective zone generated by
a TP when the temperature exceeds 300 MK. Such a condition is
typical of more massive AGB stars (4-6 M$_\odot$) \citep[see][for a recent review]{kakala} \\
The s-process elements are mainly synthesized by the \ct burning
in radiative conditions in the $^{13}$C-pocket during inter-pulse
periods. Neutron captures activated by the \ctan also lead to the
synthesis of a suitable amount of $^{15}$N, which produces
fluorine via the $^{15}$N($\alpha$,$\gamma$)$^{19}$F reaction (see
Section \ref{path}). During the first TPs, a fraction of the \ct
in the \ct-pocket may survive to the inter-pulse phase. In this
case, such \ct is engulfed into the convective shell generated by
the following TP and, thus, it burns in a convective environment
\citep{cri06}. Fluorine is also synthesized starting from the \ct
left by the H-burning shell. This latter \fluod source requires,
however, the presence of a quite large amount of C+N+O into the
H-rich
envelope.\\

In this paper we investigate fluorine nucleosynthesis in low mass
AGB stars by evaluating the effects of the variation of the
relevant nuclear reaction rates. In particular, we present the
results of an analysis based on an AGB stellar model with initial
mass M=2 M$_\odot$ and $Z=10^{-3}$. This model is representative
of low-mass AGB stars, i.e. the most promising candidates to
reproduce the majority of the observed s-process distributions in
AGB stars (see, e.g., \citealt{abia02}). Quite similar results can
be obtained for different metallicities or slightly different
masses (in the range 1.5-2.5 M$_\odot$). In this work, larger
initial stellar masses ( 4-6 M$_\odot$) are not considered since
the (eventually) produced fluorine would be efficiently destroyed
by $\alpha$ and neutron captures during TPs (see Section
\ref{path}), due to the larger temperatures attained by these
stars. Moreover, the expected surface s-process distributions of
intermediate AGB stars are characterized by low or even negative
[hs/ls] indexes and high [Rb/Sr] ratios, in striking contrast with
the observed spectra \citep{abia02}.

The aim of this work is to identify the nuclear processes whose
present uncertainties sizeably affect the theoretical predictions
and, thus, deserve further experimental investigations. A similar
study of fluorine production in AGB stars was presented almost 10
years ago by \citet{lugaro04}. With respect to that work we
extended the study to a larger number of reactions. Moreover, in
the last decade several nuclear cross sections have been measured,
or predicted, with improved precision.

In Section II we briefly present the stellar evolutionary code
used to calculate the AGB models. In Section III, the fluorine
nucleosynthesis path and the nuclear network we use are
illustrated. We also critically discuss the uncertainties
affecting all the relevant reaction rates. The results of our
analysis are illustrated and discussed in Section IV. Our
conclusions follow.

\section{Stellar models} \label{stema}

The stellar models presented in this work have been computed with
the FUNS (FUll Network Stellar) Evolutionary Code \citep[see][and
references therein]{stra06}. This code includes a full nuclear
network, with all the relevant isotopes from H up to Bi, the
heaviest element synthesized by the s-process, for a total of
nearly 500 isotopes linked by more than 1000 nuclear reactions.
The coupling of the stellar structure equations with this full
network allows to overcome some limitations related to the use of
simplified assumptions, to describe stellar phases in which the
nucleosynthesis is strongly related to the changes of the physical
properties of the stellar environment and vice versa. More details
on the nuclear network
can be found in \cite{stra06} and \cite{cri09}.\\
We make use of low temperature C-enhanced molecular opacities
\citep{cri07} to take into account the modification of the
envelope chemical composition determined by the carbon dredge up
during the TP-AGB phase. \\
Concerning mass loss, we adopt a Reimers formula with $\eta$=0.4
for the pre-AGB evolution, whilst for the AGB we fit the $\dot
M$-Period relation as determined in galactic AGB stars. We follow
a procedure similar to that described by \cite{vw93}, but taking
into account more recent infrared observations of AGB stars (see
\citealt{stra06} for the references). We find that our mass-loss
rate resembles a moderate Reimers mass loss rate for the first
part of the AGB and then switches to a stronger regime toward the
tip of the AGB, presenting however a milder slope with respect to
the rate proposed by
\cite{vw93}. \\
In our models, in order to handle the discontinuity in the
radiative gradient arising during TDU episodes, we introduce at
the base of the convective envelope an exponentially decaying
profile of convective velocities. This algorithm efficiently works
only during TDU episodes, when the H-rich (opaque) envelope
approaches the underlaying radiative He-rich layers\footnote{We
remind that the local radiative gradient (and thus the convective
velocity) is proportional to the opacity.}. As a net effect, we
obtain deeper TDUs and, as a by-product, the formation of thin \ct
pockets ($\Delta$M$<10^{-3}$M$_\odot$). The extension in mass of
the \ct pocket decreases in mass pulse after pulse, following the
natural shrinking of the whole He-intershell. Our treatment of the
discontinuity in the temperature gradient at the inner border of
the convective envelope is based on a free parameter $\beta$ which
has been calibrated in order to maximize the s-process production
in low mass AGB stars (see \citealt{cri09} for
details). \\
Of particular relevance for the present paper are
some recent updates. All the neutron capture cross sections have
been derived from the KADONIS database v 0.3 \citep{iris}, except
those of Zr isotopes \citep{ta10,ta11}, Os isotopes
\citep{mosconi10} and $^{197}$Au \citep{le11}. Concerning charged
particle reactions involved in the fluorine nucleosynthesis we
have considered available new measurements and, more in general,
we critically review the available literature (see Section
\ref{nucunce}). The list of reactions relevant for the
nucleosynthesis of $\rm^{19}F$ is reported in Table \ref{tab1}.

\section{Nuclear paths involving $^{19}$F} \label{path}

The nuclear path leading to fluorine production in low-mass AGB
stars is quite complex \citep{forestini92}. Within the
He-intershell, most of the neutrons released by the \ctan~
reaction during the inter-pulse period are captured by $^{14}$N,
via the $^{14}$N(n,p)$^{14}$C reaction and, thus, protons are
produced. Then, the $^{14}$C($\alpha$,$\gamma$)$^{18}$O reaction
synthesizes $^{18}$O which, in turn, captures the freshly
synthesized protons leading to the production of $^{15}$N via the
$^{18}$O(p,$\alpha$)$^{15}$N reaction. This process competes with
the main destruction channel of $^{15}$N, i.e. the
$^{15}$N(p,$\alpha$)$^{12}$C reaction. Later on, at the
development of the following TP, $^{15}$N captures an $\alpha$
particle producing $^{19}$F. The
$^{15}$N($\alpha$,$\gamma$)$^{19}$F reaction is therefore the
major $^{19}$F production channel. \\
A further contribution to $^{15}$N production comes from any
unburnt \ct in the \ct pocket \citep{cri09} and from the \ct left
in the H-burning ashes and engulfed in the convective zone
generated by a TP, provided a large enough C+N+O abundance in the
envelope. Such a \ct is rapidly burnt at high temperatures leading
to the synthesis of $^{15}$N through the same nuclear chain
already active in the \ct pocket. In such a case, however, an
additional contribution comes from the
$^{14}$N($\alpha$,$\gamma$)$^{18}$F($\beta^+\nu_e$)$^{18}$O
nuclear
chain.\\
A marginal contribution (less than 10\%; see, e.g.,
\citealt{ga2010}) to fluorine nucleosynthesis is given by
the $^{18}$O(n,$\gamma$)$^{19}$O$(\beta^-\bar(\nu)_e)^{19}$F nuclear chain. \\
The $^{19}$F destruction channels are the
$^{19}$F(p,$\alpha$)$^{16}$O, the $^{19}$F($\alpha$,p)$^{22}$Ne
and the $^{19}$F(n,$\gamma$)$^{20}$F reactions. In low mass AGB
stars, however, the $^{19}$F destruction is rather inefficient,
while it could be efficiently activated in higher masses
\citep{lugaro04,dora13,stra14}.

\section{Reaction rates uncertainties} \label{nucunce}

Thanks to the improvements in experimental methods and novel
techniques (e.g. Underground Nuclear Physics, Recoil Mass
Separator, Trojan Horse etc..) cross sections can be measured at
very low energies. This combined with the refinement of the
theoretical tools allowed the determination of reaction rates at
astrophysically relevant temperature without doing an
extrapolation on a wide energy range. In fact, when direct
experimental data are not available, the existence of unknown
resonance and/or the correct evaluation of resonance parameters
and/or interference pattern between different resonances may
dramatically change the cross section extrapolations. A general
discussion on techniques to extrapolate the cross sections is
beyond the aims of this paper (interested readers may refer to
\citealt{azure}). Nevertheless in the cases where the direct
experimental data are relatively close to the Gamow peak energies,
the experimental uncertainties may represent a fair description of
the reaction rate uncertainties also at relevant
energies.\\
The majority of the most relevant proton radiative captures have
been recently remeasured with improved precision extending the
data toward, and in some cases well within, the AGB relevant
energy range (see, e.g., the underground measurements of the LUNA
collaboration of $^{14}$N(p,$\gamma$)$^{15}$O
 - \citealt{imbriani05}, $^{15}$N(p,$\gamma$)$^{16}$O - \citealt{leblanc10}
and $^{17}$O(p,$\gamma$)$^{18}$F - \citealt{dileva14}). The
experimental uncertainties are relatively small also at relevant
energies, i.e. about 10-15\% for most of the processes we have
considered.\\ Also most of the (p,$\alpha$) reactions involved in
the $^{19}$F synthesis network have been re-measured in the last
10 years, reaching a comparable level of experimental precision.
For these processes the
uncertainties vary typically between 10 and 30\%. \\
In conclusion, as far as proton-induced processes are concerned,
experimental data allow reaction rate calculations at Gamow peak
energies without doing large extrapolation of the cross section.
In those cases, therefore, the reaction rate uncertainty can be
assumed to be robust.

The case of $\alpha$ capture processes is somewhat different. Due
to the much smaller penetrability, the investigated energy range
experimentally reached is, for most of the reactions, far from the
astrophysically relevant temperatures. Thus, the reaction rates
have to be estimated on the basis of cross sections extrapolated
to relevant energies by means of models often including resonance
parameters evaluated on the basis of gross estimations. As a
consequence, experimental uncertainties are much larger, i.e about
50\% in most of the cases. Moreover, it would not be unlikely that
new experimental data may produce estimates of $\alpha$ capture
reaction rates significantly different with respect to the current
ones, eventually beyond the assumed uncertainties. \\
In this view, we present the results of two different analyses. In
the first one, reaction rates are varied within the 2$\sigma$
uncertainty given by the most recent experiments available in the
literature. Then, for few selected cases, we also present a
further investigation, conduced by varying the reaction rates by
two order of magnitudes.

The experimental status for the key reactions is briefly outlined
below, including some details on the sources of the reaction rates
used in this work. The quoted uncertainties at the relevant
temperature for the fluorine nucleosynthesis represent 2$\sigma$
standard deviations.

{\bf$^{14}$N(p,$\gamma$)$^{15}$O}. In the last 10 years there have
been several experiments that measured this cross section using
both direct and indirect methods. In particular, the experiment
performed by the LUNA collaboration
\citep{imbriani05,costantini09} has reached the lowest energy,
about 70 keV, which corresponds to a stellar temperature of about
50 MK. It is worth noting that a reliable extrapolation of the
LUNA data to the solar Gamow peak requires the combination of
low-energy with high-energy data, namely  the results of the LENA
experiment at TUNL \citep{runkle05}. Additional information is
provided by experiments which used indirect methods such as the
Doppler shift attenuation method \citep{bertone01, sch08}, Coulomb
excitation \citep{yamada04} and asymptotic normalization
coefficients \cite[ANC;][]{mukhamedzhanov03}. It is worth noting
that the uncertainty on data at high energy affects the low energy
extrapolation. For this paper we use the $S$-factor recommendation
of the recent compilation Solar Fusion II \citep{SFII11}. The
uncertainty at relevant temperatures is about 10\%.

{\bf$^{15}$N(p,$\gamma$)$^{16}$O}. The low energy cross section of
this process is determined by the presence of two broad resonances
and by their interference. Recent studies that used indirect
methods \citep{mukhamedzhanov08, lacognata09} suggested a cross
section significantly lower than previously estimated. A direct
experiment \citep{leblanc10, Imbriani2012} performed in a wide
energy window, confirmed that the cross section is about a factor
of 2 lower than previously thought. This translates into a
significantly lower rate at AGB temperatures. In the present work
we use the rate, and the corresponding uncertainty, presented in
\cite{leblanc10}. The uncertainty at relevant temperatures is
about 10\%.

{\bf$^{17}$O(p,$\gamma$)$^{18}$F}. The cross section of this
reaction below $E_{\rm cm} \simeq400\rm\,keV$ is determined by 2
narrow resonances, together with the tails of 2 broad higher
energy resonances and the direct capture (DC) component. After the
pioneering work of \citet{rolfs73} in the seventies, in the last
decade several studies \citep{Fox05, Chafa07, Newton10, DRAGON12,
Kontos12, scott12} have determined the reaction rate with
improving precision. Nevertheless, at the temperatures relevant
for AGB nucleosynthesis, the reaction rate is dominated by the
lowest energy resonance, $E_{\rm cm}=65\,$keV, that is too weak to
be directly measured with the current experiment possibilities.
For the present work we use the reaction rate determination by
\cite{dile14} based on the measurement of the LUNA collaboration.
The uncertainty at relevant temperatures is about 15\%.

{\bf$^{18}$O(p,$\gamma$)$^{19}$F}. Also in this case the presence
of several low energy states influences the determination of the
cross section. In particular, the $E_{\rm cm} = 150\,$keV broad
resonance and the direct capture dominate the reaction rate at the
astrophysically relevant temperature for AGB nucleosynthesis.
\citet{wiescher80} and \citet{vogelaar90} have directly determined
the magnitude of all the relevant states, down to $E_{\rm
cm}=89\rm\,keV$. For the present work we adopt the reaction rate
given in the recent compilation by \cite{iliadis2010b}. The
uncertainty at relevant temperatures is about 10\%.

{\bf$^{15}$N(p,$\alpha$)$^{12}$C}. The reaction rate of this
process, at the temperatures of interest for this study, is mostly
determined by a resonance at $E_{\rm cm}\simeq100\rm\,keV$. The
cross section was directly determined by \cite{redder82}. More
recently, this reaction has been investigated again using the
indirect approach of the Trojan Horse Method
\citep[THM;][]{lacognata07}, yielding results similar to
\cite{redder82}, for both the central value and the uncertainty.
For the present work we use the reaction rate reported in the
NACRE compilation \cite{nacre}. The uncertainty at relevant
temperatures is about 10\%.

{\bf$^{17}$O(p,$\alpha$)$^{14}$N}. Several experiments
\citep{Chafa07, Newton10, Moazen07}  determined the magnitude of
most of the several resonances that influence the rate of this
reaction. As for the $^{17}$O(p,$\gamma$)$^{18}$F the presence of
the $E_{\rm cm} = 65\rm\,keV$ resonance, directly at the
astrophysically relevant energy, makes the reaction rate
determination difficult and the uncertainty is correspondingly
large. For the present work we adopt the value suggested by
\cite{iliadis2010b}. The uncertainty at relevant temperatures is
about 20\%.

{\bf$^{18}$O(p,$\alpha$)$^{15}$N}. The low energy determination of
this cross section is complicated by the tail of high energy broad
resonances and by the presence of several low energy states
\citep{Lorentz-Wirzba78}. Recently \citet{lacognata10} have
remeasured with the THM this cross section, strongly reducing the
corresponding uncertainty. For the present calculations we use the
reaction rate provided by \cite{iliadis2010b}. The uncertainty at
relevant temperatures is up to 30\%.

{\bf$^{19}$F(p,$\alpha$)$^{16}$O}. The situation is very similar
to the previous case. The presence of very low energy resonances
makes direct determination of the cross section very difficult. An
indirect experiment using the THM \citep{lacognata11} has observed
the presence of resonances not seen before in this process at
energies corresponding to typical AGB temperatures, thus implying
a significant increase of the reaction rate. Thus, the reaction
rate proposed by \cite{lacognata11} is used in our calculations.
The uncertainty at relevant temperatures is about 30\%.

{\bf$^{14}$C($\alpha$,$\gamma$)$^{18}$O}. The determination of
this cross section is not easy, because of the difficulty of
$^{14}$C targets or beams production. Few direct experiments have
been performed  to determine this cross section \citep{goerres92}.
Recently the cross section was determined through indirect
techniques by \cite{johnson09}. A better knowledge of several
resonances was achieved with a significant improvement in the
uncertainty in the reaction rate at temperatures lower than
typical for AGBs. In the temperature window relevant for our study
there is a general agreement between the reaction rates provided
by \cite{goerres92} and \cite{johnson09}. In our calculations we
used the reaction rate given by \cite{lugaro04}, which is based on
the \cite{goerres92} data with the exception of the spectroscopic
factor of the 6.2 MeV state, for which they used a higher value.
The uncertainty at relevant temperatures is about 50\%.

{\bf$^{18}$O($\alpha$,$\gamma$)$^{22}$Ne}. The determination of
this reaction rate is particularly complex and several experiments
have been done to collect all the necessary information. The
determination is relatively uncertain at lower energy. We followed
the prescriptions by \cite{iliadis2010b}. The uncertainty at
relevant temperatures is about 50\%.

{\bf$^{19}$F($\alpha$,p)$^{20}$Ne}. The only experiment available
has been done by \cite{ugalde08}. These authors have measured this
cross section, using an $\alpha$ beam and a F solid target between
E$_{cm}$ = 700 and 1600 keV, still far from the relevant
astrophysical energy. The uncertainty at relevant temperatures is
up to 50\%.

{\bf$^{13}$C($\alpha$,n)$^{16}$O}. This reaction is the major
neutron source in low mass AGB star and releases neutrons in
radiative conditions during the inter-pulse period at temperatures
T $\sim 10^8\,$K. The $^{13}$C($\alpha$, n)$^{16}$O reaction has
been studied over a wide energy range, but none of the existing
experimental investigations has fully covered the relevant
astrophysical energies. The present low energy limit in a direct
experiment is $E_{\rm cm}$ = 220 keV, while the astrophysical
relevant energy is between $\sim$ 150 and 250 keV. This means that
the cross section at the typical AGB temperatures has to be
extrapolated from the data through model predictions. For our
calculation we have used the reaction rate derived by
\citet{heil08}. The uncertainty at relevant temperatures is up to
50\%.
%

{$\rm \mathbf{^{14}N(\alpha,\gamma)^{18}F}$.} The rate of the $\rm
^{14}N(\alpha,\gamma)^{18}F$ reaction is dominated by the
contribution of a $J^\pi=1^-$ resonance at $E_{\rm cm} =
572\rm\,keV$ for temperatures between $0.1\,$GK and $0.5\,$GK.
For higher temperatures the contributions of resonances at higher
energies ($E_{\rm cm} = 883, 1088, 1189$ and $1258\rm\,keV$)
become more important. Below $0.1\,$GK additional contributions
are possible from the low-energy tail of the $445\rm\,keV$
resonance, the DC component, and the $J^\pi=4^+$ resonance at
$237\rm\,keV$. The lowest directly measured resonance is the one
at $445\rm\,keV$, which has been measured together with the
resonance at $1136\rm\,keV$ \citep{Goerres2000}.

\noindent The strength of the $E_{\rm cm}=237\rm\,keV$ resonance
has never been measured. It has been argued \citep{Goerres2000}
that it is extremely small
($\omega\gamma\sim1.5\cdot10^{-18}\rm\,keV$), far below present
experimental measurement possibilities.

\noindent The DC for this reaction is only theoretically
estimated. It is supposed to be $\sim0.7-0.8\rm\,keV$b, as
calculated from spectroscopic factors using some crude
approximations \citep[see comments in][]{iliadis2010b}.
The uncertainty at relevant temperatures is up to 20\%.

{$\rm\mathbf{^{15}N(\alpha,\gamma)^{19}F}$}. The rate of the $\rm
^{15}N(\alpha,\gamma)^{19}F$ reaction is dominated by resonance
contributions of several low-lying states in $\rm^{19}F$.
\citet{Wilmes2002} measured all the resonance strengths in the
energy window $0.6\rm\,MeV$ to $2.7\rm\,MeV$. Nevertheless at
astrophysical relevant energies contributions from the $E_{\rm
cm}$ = 364 keV, 536 keV and 542 keV resonances are
important.\\
\begin{figure}[h]
\includegraphics[width=.9\columnwidth]{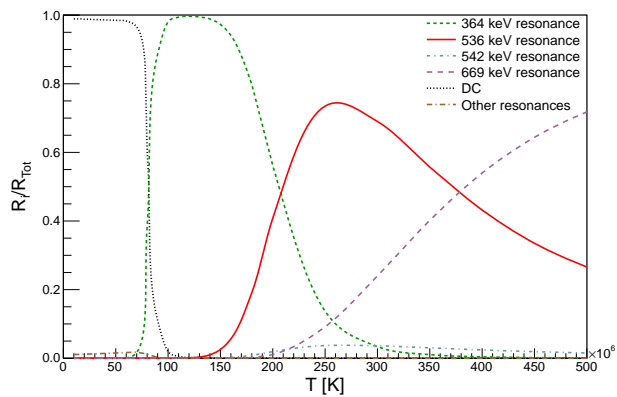}
\caption{\label{fig:rate15Nag} Fractional contribution of the
different resonances and the DC to the
$\rm{^{15}N(\alpha,\gamma)^{19}F}$ reaction rate.}
\end{figure}
The strength of the $364\rm\,keV$ resonance has been measured
indirectly \cite{DeOliveira1996} using the $^{15}{\rm N}(^{7}{\rm
Li},t)\rm ^{19}F$ reaction at $28\,$MeV, the uncertainty
associated with the model used to derive the results is estimated
to be a factor of 2. This resonance directly influences the
determination of the reaction rate for temperatures lower than
$200\rm\,MK$. Thus, a direct measurement of its strength is of
great importance. In Figure \ref{fig:rate15Nag} we show the
contributions of the resonances and the DC to the total rate
versus the temperature for the
$\mathrm{^{15}N(\alpha,\gamma)^{19}F}$. The uncertainty ranges
from 15\% at 250 MK, up to a factor 2 below 180 MK (see Table
\ref{tab2}).

Considering the minor contribution of the
$^{18}$O(n,$\gamma$)$^{19}$O reaction to \fluod production we
exclude this reaction from our analysis. Similarly, we do not take
into account variations of the $^{19}$F(n,$\gamma$)$^{20}$F, since
in low mass stars this reaction is not efficiently activated.

A list of the studied reaction rates is provided in Table
\ref{tab2} (column 1). Upper and lower percentage cross section
uncertainties (2$\sigma$) are reported at T=$1\times 10^8$ K and
T=$2.5\times 10^8$ K (column 2-3 and 4-5). These are the typical
temperatures of the nucleosynthesis occurring in the \ct pockets
and in the convective shells generated by thermal pulses,
respectively.

\section{Results}

As a first step, we compute a model by using the new set of
reaction rates reported in Table \ref{tab1} (hereinafter ST case).
We select a model with initial mass M=2 M$_\odot$ and $Z=10^{-3}$
([Fe/H]=-1.15), considered as representative of low mass AGB
stars. The initial He abundance is set to 0.245; the distribution
of metals is assumed to be solar scaled and taken from
\cite{lo03}. No $\alpha$-element enhancements are considered. The
mixing length parameter ($\alpha_{m.l.}$) has been calibrated by
computing a standard Solar Model and its value is set to 2.1
\citep{pi07}. The evolution of this model has been stopped after
the last TDU, when the core has a mass of 0.686 M$_\odot$ and the
mass of the remaining convective envelope is smaller than $7\times
10^{-3}$ M$_\odot$. During its evolution, this model experience 20
thermal pulses, 19 of which followed by TDU. For
more details we refer to \cite{cri09}. \\
With respect to a model computed with the previous version of the
network, we obtain a reduction of the surface [F/Fe] of 29\% (and
a corresponding reduction of the [F/$\langle s\rangle$] of 22\%).
This result is mainly ascribed to the strongly reduced rate (about
a factor 1000 at relevant temperatures) of the
$^{14}$C($\alpha$,$\gamma$)$^{18}$O reaction \citep{lugaro04} with
respect to the rate adopted in our previous calculations
\citep{cf88}. Such a low rate limits the local increase of
$^{18}$O. Later on, part of this $^{14}$C is dredged up to the
surface and it decays to $^{14}$N, thus not directly contributing
to fluorine
nucleosynthesis within the He-intershell. \\

Then, we assume as a reference case the model computed with such a
new nuclear network (hereafter ST case) and we vary one at a time
all the reactions reported in Table \ref{tab2} in both directions
(upper and lower limits). For each varied reaction, we compute the
evolution of the star from the end of the core He-burning phase up
to the tip of the AGB phase.

\subsection{2$\sigma$ analysis}

We firstly studied the effects on fluorine nucleosynthesis due to
variations of the relevant reaction rates by changing them within
2$\sigma$ values, as derived from the extant literature.

Results are shown in Table \ref{tab2}, where we tabulate the final
(i.e. after the last TDU) surface [F/Fe] and [F/$\rm\langle
s\rangle$] percentage differences with respect to our ST case. For
both quantities we provide variations corresponding to upper and
lower rates. From the values reported in Table \ref{tab2} it comes
out that the variations of key reaction rates within $2\sigma$
have little effect on the final fluorine surface abundances and
[F/$\rm\langle s\rangle$] ratios. Largest variations are found for
the $^{14}$N(p,$\gamma$)$^{15}$O, the
$^{19}$F($\alpha$,p)$^{22}$Ne and the $^{13}$C($\alpha$,n)$^{16}$O
reactions. However, such differences in the final surface fluorine
abundance are definitely lower than the errors
currently affecting spectroscopic observations (at least a factor 2). \\

\subsection{Exploration above current experimental uncertainties} \label{secnoc}

In order to better evaluate the sensitivity of $^{19}$F
nucleosynthesis to larger variations of selected reaction rates,
we run some additional tests by varying the cross section of
$\alpha$ capture processes well beyond the currently quoted
uncertainties. The computed tests are listed in Table \ref{tab3},
where we report:\\
\begin{itemize} \item{scaling factors $sf$
(i.e. the ratio of the modified rate with respect to the reference
case);} \item{corresponding final surface $^{19}$F ratios with
respect to the reference case (R$({^{19}{\rm F}})=^{19}{\rm
F}_{test}/^{19}{\rm F}_{ST}$);} \item{corresponding final surface
F/$<s>$ ratios with respect to the reference case (R$({{\rm
F}/<{\rm s}>})=({\rm F}/<{\rm s}>)_{test}/({\rm F}/<{\rm
s}>)_{ST})$.} \end{itemize} Among the studied rates, the
$^{14}$C($\alpha$,$\gamma$)$^{18}$O and the
$^{18}$O($\alpha$,$\gamma$)$^{22}$Ne show the lowest variations.
Large changes of the $^{14}$N($\alpha$,$\gamma$)$^{18}$F and the
$^{13}$C($\alpha$,n)$^{16}$O reactions imply small, but
appreciable, differences in the final fluorine surface abundances.
The remaining two rates, {\it i.e.} the
$^{15}$N($\alpha$,$\gamma$)$^{19}$F and the
$^{19}$F($\alpha$,p)$^{22}$Ne are able to reduce by a factor of 10
the surface $^{19}$F abundance and, thus, deserve a more careful
analysis.

\subsubsection{The revisited $^{15}$N($\alpha,\gamma$)$^{19}$F}
\label{revn15}

Interestingly enough, the model with a strongly reduced
$^{15}$N($\alpha$,$\gamma$)$^{19}$F roughly matches the observed
[F/$<s>$] ratios found at low metallicity by \cite{abia11}. Note
that, as we already remarked, the
$^{15}$N($\alpha$,$\gamma$)$^{19}$F rate is strongly influenced by
the presence of the $364\rm\,keV$ resonance, whose intensity is
affected by a very large uncertainty. The contribution from the DC
component is also difficult to evaluate at the present status of
the experimental knowledge. Therefore, although very unlikely, the
explored scenario is not impossible from a nuclear point of view.
Nevertheless, a very low value of the
$^{15}$N($\alpha$,$\gamma$)$^{19}$F reaction would not allow AGB
stars to synthesize fluorine at solar-like metallicities, in
contradiction with observations \citep{abia10}. As already
recalled, however, preliminary studies based on a recent reduction
of the adopted excitation energies of HF molecular transitions
seem to lead to a consistent decrease of the observed fluorine
surface abundances in solar-like metalliticy AGB stars (Abia et
al., in preparation).
\\
A revision of this reaction rate looks therefore mandatory, in
order to minimize the uncertainty on \fluod nucleosynthesis
related to this nuclear process. We will clarify this issue in the
near future, when we will directly measure the strength of the
$364\rm\,keV$ resonance and the intensity of the DC component,
using the ERNA recoil separator
\citep{Schuermann2005,DiLeva2009,DiLeva2012}.

\subsubsection{The revisited $^{19}$F($\alpha$,p)$^{22}$Ne}
\label{revf19}

The only other $\alpha$ capture process able to significantly
change the surface fluorine abundances in low mass AGB stars is
the $^{19}$F($\alpha$,p)$^{22}$Ne. In state-of-the-art modelling,
this reaction rate is expected to efficiently work in stars with
larger masses (M$\geq 4-5 $M$_\odot$ \footnote{Note that, as
already highlighted in the Introduction, this class of objects
cannot reproduce the observed AGB s-process distributions.}).
However, a deep revision of this rate cannot be excluded from
first principles. In fact, the $^{19}$F($\alpha$,p)$^{22}$Ne cross
section is characterized by many broad resonances tailing into the
low-energy range. Possible additional resonance contributions in
that excitation range were predicted, on the basis of $^{23}$Na
compound nuclear level structure data. As discussed in Section
\ref{nucunce}, experimental data are still far from the relevant
astrophysical energies. The uncertainty is mostly due to the
different energy dependence of the cross section predicted by
different models. We intend to clarify this point by realizing a
direct measurement of this cross section at energy lower than
E$_{cm}$ =660 keV, using a fluorine beam impinging into an
$\alpha$ jet gas target surrounded by particle detectors.

\section{Conclusions}

In this paper we investigated the effects that nuclear reaction
rates have on the fluorine production in low mass AGB stars. \\
We found minor variations in the final surface \fluod abundance
(less than 10\%) when varying the rates within 2$\sigma$ standard
deviations. However, some $\alpha$ capture reactions may change
well beyond the current accepted uncertainties, due to the lack of
knowledge about the strengths of nuclear resonances at
astrophysical relevant energies. Thus, we investigated further
possible, although unlikely, scenarios in which we varied some
$\alpha$ capture rates by a factor 100. We found that the largest
variations are obtained by reducing the
$^{15}$N($\alpha$,$\gamma$)$^{19}$F rate or by increasing the
$^{19}$F($\alpha$,p)$^{22}$Ne rate. Both reactions have poorly
known resonances at astrophysical energies. We plan to
experimentally investigate both cross sections in the future
directly measuring the strengths of their resonances.

\begin{acknowledgements} This work was supported by Italian Grants RBFR08549F-002 (FIRB
2008 program), PRIN-INAF 2011 project "Multiple populations in
Globular Clusters: their role in the Galaxy assembly", PRIN-MIUR
2012 "Nucleosynthesis in AGB stars: an integrated approach"
project (20128PCN59) and from Spanish grants AYA2008-04211-C02-02
and AYA-2011-22460.
\end{acknowledgements}

\bibliographystyle{aa} \bibliography{ms}

\newpage
\begin{center}
\begin{table*}
\caption{\label{tab1} Sources of the reaction rates relevant for
fluorine nucleosynthesis: "Old source" list corresponds to the
network used in \citealt{stra06} and \citealt{ cri09}; "New
source" list refers to the present work.}
\begin{tabular}{lcc}
Reaction rate &  Old Source & New Source \\
\hline1
Proton captures & & \\
\hline
$^{14}$N(p,$\gamma$)$^{15}$O & \cite{formicola} & \cite{SFII11}\\
$^{15}$N(p,$\gamma$)$^{16}$O & \cite{nacre} &  \cite{leblanc10} \\
$^{17}$O(p,$\gamma$)$^{18}$F & \cite{nacre} & \cite{scott12} \\
$^{18}$O(p,$\gamma$)$^{19}$F & \cite{nacre} & \cite{iliadis2010b}\\
$^{15}$N(p,$\alpha$)$^{12}$C & \cite{nacre} & \cite{nacre} \\
$^{17}$O(p,$\alpha$)$^{14}$N & \cite{nacre} & \cite{iliadis2010b} \\
$^{18}$O(p,$\alpha$)$^{15}$N & \cite{nacre} & \cite{iliadis2010b}\\
$^{19}$F(p,$\alpha$)$^{16}$O & \cite{nacre} & \cite{lacognata11} \\
\hline
$\alpha$ captures & & \\
\hline
$^{14}$C($\alpha$,$\gamma$)$^{18}$O  & \cite{cf88} & \cite{lugaro04} \\
$^{14}$N($\alpha$,$\gamma$)$^{18}$F  & \cite{goerres00} & \cite{iliadis2010b} \\
$^{15}$N($\alpha$,$\gamma$)$^{19}$F  & \cite{nacre} &  \cite{iliadis2010b}\\
$^{18}$O($\alpha$,$\gamma$)$^{22}$Ne & \cite{giesen94} &  \cite{iliadis2010b}\\
$^{19}$F($\alpha$,p)$^{22}$Ne & \cite{ugalde05} & \cite{ugalde08}\\
$^{13}$C($\alpha$,n)$^{16}$O & \cite{drotleff} &\cite{heil08}\\
\end{tabular}
\end{table*}
\end{center}

\newpage
\begin{center}
\begin{table*}
\caption{$2\sigma$ percentage cross section upper and lower
uncertainties at T=$1\times 10^8$ K and T=$2.5\times 10^8$ K and
corresponding percentage fluorine surface variations. See text for
details.\label{tab2}}
\begin{tabular}{lcccccccc}
 &\multicolumn{2}{c}{$2\sigma$  $(T_8=1)$ }& \multicolumn{2}{c}{$2\sigma$  $(T_8=2.5)$ }& \multicolumn{2}{c}{$\Delta$ [F/Fe] (\% var.)}& \multicolumn{2}{c}{$\Delta$ [F/$\langle s\rangle$] (\% var.)} \\
Reaction rate & Upper &Lower & Upper & Lower & Upper& Lower & Upper & Lower \\
\hline
$^{14}$N(p,$\gamma$)$^{15}$O         & 10&10 &8 & 8& -3 & +5 & -3 & +3 \\
$^{15}$N(p,$\gamma$)$^{16}$O         & 15& 15& 15& 15& -1 & -2 & -3 & -2 \\
$^{17}$O(p,$\gamma$)$^{18}$F         & 15& 15& 20& 20&  0 & -2 & -3 &  0 \\
$^{18}$O(p,$\gamma$)$^{19}$F         & 30& 30& 30& 30& -2 & -3 & -1 & -3 \\
$^{15}$N(p,$\alpha$)$^{12}$C         &20 &20 &15 &15 & -3 & +1 & -3 & -3 \\
$^{17}$O(p,$\alpha$)$^{14}$C         & 15& 15& 6& 6& -2 & -2 & -1 &  0 \\
$^{18}$O(p,$\alpha$)$^{15}$N         &8 &8 & 8& 8& +1 & -2 & +3 & -1 \\
$^{19}$F(p,$\alpha$)$^{16}$O         & 35& 35& 35& 35&  0 & -1 & -4 & -4 \\
$^{14}$C($\alpha$,$\gamma$)$^{18}$O  & 100&84 &100 &62 & -2 &  0 & -3 & -2 \\
$^{14}$N($\alpha$,$\gamma$)$^{18}$F  &20 &20 &10 &10 & -1 & -1 & +3 & -1 \\
$^{15}$N($\alpha$,$\gamma$)$^{19}$F  &100 &50 &15 &15 & -3 & -2 &  0 & +5 \\
$^{18}$O($\alpha$,$\gamma$)$^{22}$Ne & 70& 50& 70& 50& -3 & +1 & -4 & -5 \\
$^{19}$F($\alpha$,p)$^{22}$Ne        &100 & 100& 50& 50& -5 & +2 & -2 & +4 \\
$^{13}$C($\alpha$,n)$^{16}$O         & 25& 25& 25& 25& -3 & +7 & -1 & +3  \\
\hline \multicolumn{9}{l}{$T_8$: temperature in units of $10^8$
K}.
\end{tabular}
\end{table*}
\end{center}

\newpage
\begin{center}
\begin{table*}
\caption{Scaling factors $sf$ of the computed tests with the
corresponding $^{19}$F and F/$<s>$ surface ratios with respect to
the reference case. See text for details.\label{tab3}}
\begin{tabular}{lcccc}
Reaction rate & $sf$ & R$({^{19}{\rm F}})$ & R$({{\rm F}/<{\rm s}>})$ \\
\hline
$^{13}$C($\alpha$,n)$^{16}$O         & 0.01 & 4.70 &  2.80  \\
$^{13}$C($\alpha$,n)$^{16}$O         & 100  & 0.62 &  0.67 \\
$^{14}$C($\alpha$,$\gamma$)$^{18}$O  & 0.01 & 1.03 & 1.59    \\
$^{14}$C($\alpha$,$\gamma$)$^{18}$O  & 100 &  1.04 & 1.61   \\
$^{14}$N($\alpha$,$\gamma$)$^{18}$F  & 0.01 & 3.03 & 5.14  \\
$^{14}$N($\alpha$,$\gamma$)$^{18}$F  & 100  & 0.64 & 1.10      \\
$^{15}$N($\alpha$,$\gamma$)$^{19}$F  & 0.01 & 0.11 & 0.12  \\
$^{15}$N($\alpha$,$\gamma$)$^{19}$F  & 100 &  0.96 & 1.50     \\
$^{18}$O($\alpha$,$\gamma$)$^{22}$Ne & 0.01 & 2.21 & 2.01  \\
$^{18}$O($\alpha$,$\gamma$)$^{22}$Ne & 100 &  0.52 & 0.52 \\
$^{19}$F($\alpha$,p)$^{22}$Ne        & 0.01 & 1.05 & 1.19   \\
$^{19}$F($\alpha$,p)$^{22}$Ne        & 100  & 0.08 & 0.14  \\
\end{tabular}
\end{table*}
\end{center}

\end{document}